\documentclass[mathleft
]{an}
\usepackage{graphicx}

\usepackage{times}
\begin{document}

\Pagespan{1}{}
\Yearpublication{2014}%
\Yearsubmission{2014}%
\Month{11}%
\Volume{999}%
\Issue{88}%

\title{
New spectroscopic and polarimetric observations of the A0 supergiant HD\,92207\thanks
{Based on observations collected with the CORALIE echelle spectrograph mounted on the 1.2-m Swiss telescope at
 La Silla Observatory, data obtained at the European Southern Observatory (ESO 
Prg.\ 092.D-0209(A), and data obtained from the ESO Science Archive Facility under 
request MSCHOELLER 102067).}}

\date{Received date / Accepted date}

\author{
{S. Hubrig\inst{1}\fnmsep\thanks{Corresponding author: \email{shubrig@aip.de}}}
\and
A.~F.~Kholtygin\inst{2}
\and
M.~Sch\"oller\inst{3}
\and
R.~I.~Anderson\inst{4,5}
\and
S.~Saesen\inst{4}
\and
J.~F.~Gonz\'alez\inst{6}
\and
I.~Ilyin\inst{1}
\and
M.~Briquet\inst{7}\fnmsep\thanks{F.R.S.-FNRS Postdoctoral Researcher, Belgium}
}

\titlerunning{Magnetic field of HD\,92207}

\authorrunning{S.\ Hubrig et al.}

\institute{
Leibniz-Institut f\"ur Astrophysik Potsdam (AIP), An der Sternwarte 16, 14482 Potsdam, Germany
\and
Astronomical Institute, Saint-Petersburg State University, Universitskii pr.~28, 198504 Saint-Petersburg, Russia 
\and
European Southern Observatory, Karl-Schwarzschild-Str.~2, 85748 Garching, Germany
\and
Geneva Observatory, Geneva University, Chemin des Maillettes 51, 1290 Sauverny, Switzerland
\and
Department of Physics and Astronomy, Johns Hopkins University, Baltimore, MD 21218, USA
\and
Instituto de Ciencias Astronomicas, de la Tierra, y del Espacio (ICATE), 5400 San Juan, Argentina 
\and
Institut d'Astrophysique et de G\'eophysique, Universit\'e de Li\`ege, All\'ee du 6 Ao\^ut 17, Sart-Tilman, 
B\^at.\ B5C, 4000, Li\`ege, Belgium
}


\abstract{
Our recent search for the presence of a magnetic field in 
the bright early A-type supergiant HD\,92207 using FORS\,2 in spectropolarimetric mode 
revealed the presence of a longitudinal magnetic field of the order of a few 
hundred Gauss. However, the definite 
confirmation of the magnetic nature of this object remained pending due to the detection of short-term spectral 
variability probably
affecting the position of line profiles in left- and right-hand polarized spectra. 
We present new magnetic field measurements of HD\,92207 obtained on three different epochs in 2013 and 2014 using FORS\,2 in 
spectropolarimetric mode. A 3$\sigma$ detection of the mean longitudinal magnetic field using the entire spectrum, 
$\left<B_{\rm z}\right>_{\rm all}=104\pm34$\,G, was achieved in observations obtained in 2014 January.
At this epoch, the position of the spectral lines appeared stable. Our analysis of spectral line shapes
recorded in opposite circularly polarized light, i.e.\ in light with opposite sense of rotation, reveals
that line profiles in the light polarized in a certain direction appear slightly split. The mechanism causing
such a behaviour in the circularly polarized light is currently unknown.
Trying to settle the issue of short-term variability, we searched for changes in the spectral 
line profiles on a time scale of 8--10\,min using 
HARPS polarimetric spectra and on a time scale of 3--4\,min using time series obtained with the CORALIE spectrograph.
No significant variability was detected on these time scales during the epochs studied.
}

\maketitle

\section{Introduction}
\label{sect:intro}

 Our recent search for the presence of a magnetic field in 
the visually brightest early A-type supergiant HD\,92207, using FORS\,2 in spectropolarimetric mode,
revealed the presence of a weak mean longitudinal magnetic field of the order of a few 
hundred Gauss (Hubrig et al.\ \cite{Hubrig2012}).
This target has been monitored for several years in 
the $uvby$-Str\"omgren system by Sterken (\cite{Sterken1983}) and spectroscopically by 
Kaufer et al.\ (\cite{Kaufer1996,Kaufer1997}), 
who found cyclical changes of the brightness and substantial profile changes for metal lines and
H$\alpha$, and suggested that the observed photometric and H$\alpha$ line variations are the result 
of a corotating structure in the wind, which they considered to be in the star's equatorial plane. 
Furthermore, their study of the line profile variations revealed clear pulsation-like structures, 
indicating the presence of non-radial pulsations (NRPs) with a period of 27\,days, while the stellar 
rotation period is of the order of several months.
Ignace et al.\ (\cite{Ignace2009}) measured linear polarization 
in the spectra of this star on seven different nights, 
spanning approximately three months in time. For the continuum polarization, the 
spiral-shaped wind density enhancement in the equatorial plane of the star suggested by 
Kaufer et al.\ (\cite{Kaufer1996}) was explored.
Importantly, the authors reported that the polarization across 
the H$\alpha$ line on any given night is typically different from the degree and position angle 
of the polarization in the continuum. These night-to-night variations in the H$\alpha$ polarization 
are hard to understand in terms of the spiral structure that was considered for the continuum polarization. 
The determination of the atmospheric parameters was carried out by Przybilla et al. (\cite{Prz2006}), who also 
established elemental abundances for over 20 chemical species.

In the subsequent spectropolarimetric study on this target, Hubrig et al.\ (\cite{Hubrig2014})
presented a careful inspection of the FORS\,2 spectra used for the 
magnetic field determination in HD\,92207 and reported the detection of short-term spectral variability 
on a time-scale of minutes or fractions of minutes. In the framework of that study, the authors 
convincingly showed the absence of  
instrument instabilities or flexures, presenting significantly stable spectral line profiles for a number of
other stars observed quasi-simultaneously and nearby in the sky to HD\,92207. 
However, with the available data, it was not possible to decide
whether the discovered variations in line profiles of HD\,92207 were of periodic or stochastic nature. 
In any case, given the size of the supergiant, 
it was clear that the variability could not be referred to coherent line variations across the 
entire surface on such short time-scales. Hubrig et al.\ (\cite{Hubrig2014}) estimated the length of the rotation period 
as $P_{\rm rot}\le235\pm36$\,d.

The question how pulsations affect the magnetic field measurements is not yet solved in spite of 
the fact that the number of studies of pulsating $\beta$~Cephei and slowly pulsating B (SPB) stars is 
gradually increasing.
Already in \cite{Schnerr2006}, Schnerr et al.\ discussed the
influence of pulsations on the analysis of the magnetic field strength in the $\beta$~Cephei star $\nu$~Eri
in MUSICOS spectra and tried to model the signatures found in Stokes~$V$ and $N$ spectra. 
Although the authors wrote that using some modeling they were able to quantitatively establish
the influence of pulsations on the magnetic field determination, they still detected profiles in
Stokes~$N$ and $V$ that were the result of the combined effects of the pulsations and the inaccuracies
in wavelength calibration that were not removed by their imperfect modeling of these effects.

Clearly, it is not possible to use the low-resolution FORS\,2 spectra to model the effect of
pulsations on the magnetic field measurements, and the potential of high-resolution spectropolarimetric
observations should be used in the search of short-term variations (e.g.\ Hubrig et al.\ \cite{Hubrig2011a}).
On the other hand, it is possible that the degree with which pulsations impact the magnetic field 
measurements is time-dependent, e.g., if the line variability on short time scales were more or less pronounced  
at certain rotation phases. 
To investigate this possibility, we re-observed HD\,92207 with FORS\,2 
on three different epochs in December 2013, January 2014, and February 2014 in the 
framework of the ESO Programme 092.D-0209(A).

\begin{figure*}
\centering
\includegraphics[width=0.32\textwidth]{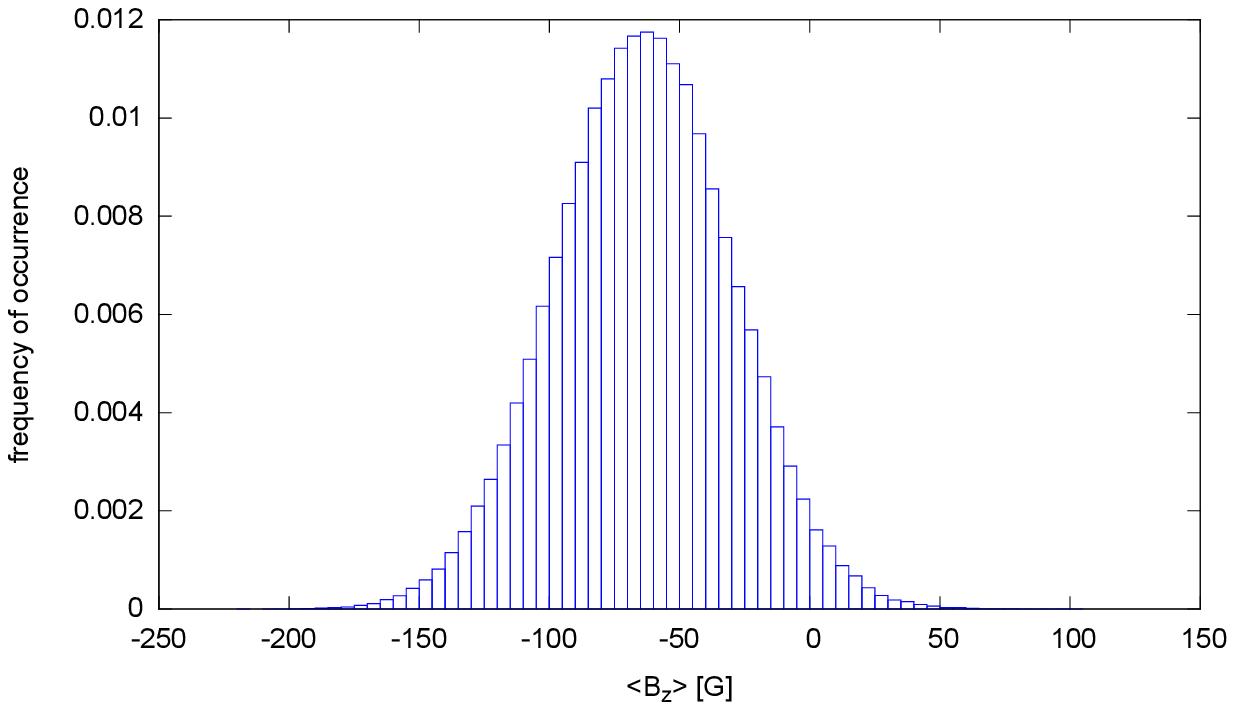}
\includegraphics[width=0.32\textwidth]{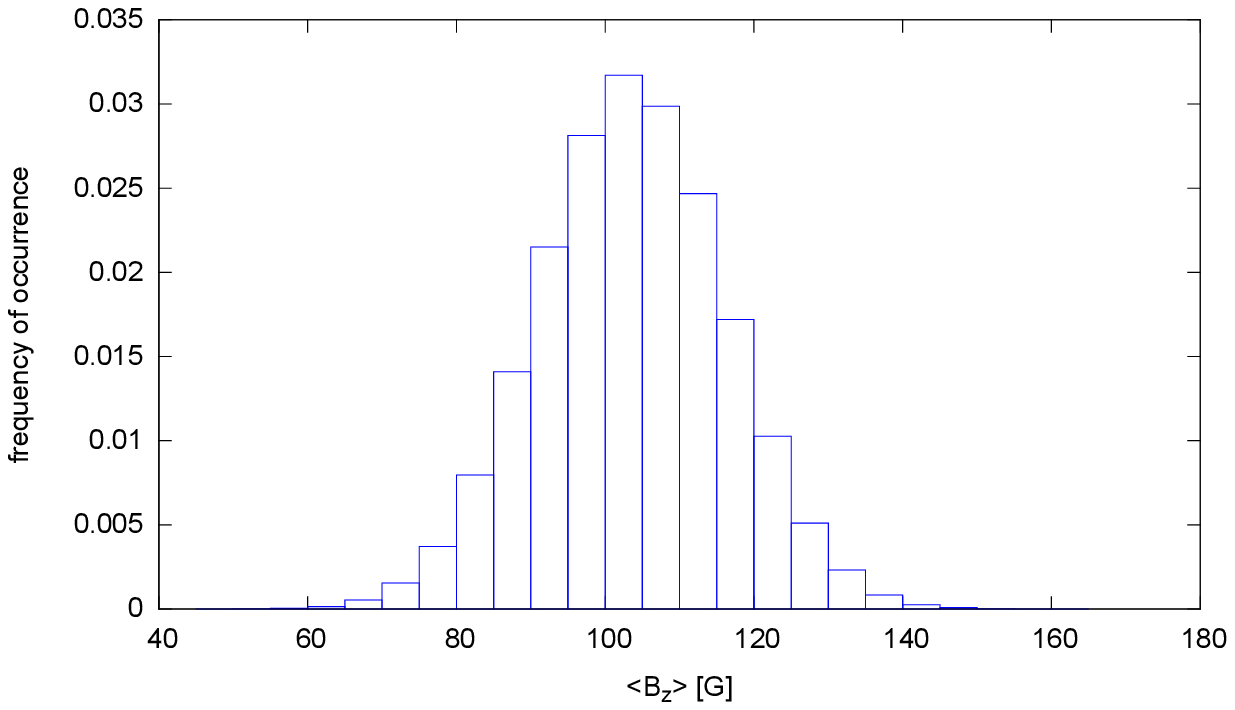}
\includegraphics[width=0.32\textwidth]{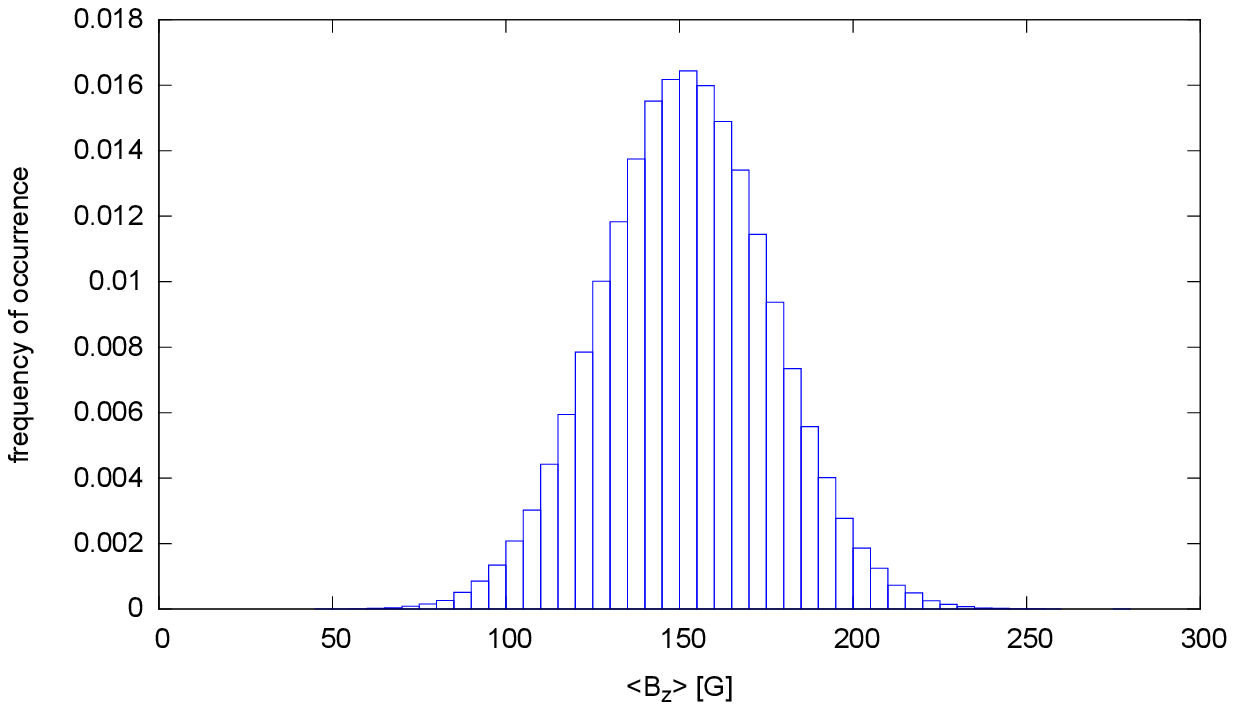}
\caption{
Distributions from our Monte Carlo bootstrapping tests for the data sets of
HD\,92207 from 2013 December (left panel), 2014 January (middle panel), and 2014 February (right panel),
using the entire spectra.
The widths of these profiles are used to determine the errors listed in Table~\ref{tab:log_measfors}.
}
\label{fig:boot}
\end{figure*}

In this study, we discuss the new magnetic field measurements of HD\,92207. In addition, we present 
the results of our analysis of HARPS archival spectropolarimetric daat and of our search for 
short-term variability 
using time series with the CORALIE spectrograph installed at the Swiss 1.2-m Euler telescope
on La Silla Observatory, Chile.

\section{Measurements of the magnetic field using FORS\,2 observations}

\subsection{Observations and data reduction}
\label{sect:fors_obs}

Three new spectropolarimetric observations of HD\,92207 were carried
out in service mode at the European
Southern Observatory with FORS\,2 mounted on the 8-m Antu telescope of the
VLT from 2013 December 4 to 2014 February 14.
This multi-mode instrument is equipped with
polarization analyzing optics, comprising super-achromatic half-wave and quarter-wave 
phase retarder plates, and a Wollaston prism with a beam divergence of 22$\arcsec$  in 
standard resolution mode.

Polarimetric spectra were obtained with the GRISM~600B and 
the narrowest slit width of 0$\farcs$4 to achieve 
a spectral resolving power of $R\sim2000$. The use of the mosaic detector 
with a  pixel size of 15\,$\mu$m allowed us to cover a
spectral range from 3250 to 6215\,\AA{}, which includes all hydrogen Balmer lines 
from H$\beta$ to the Balmer jump. 

From the raw FORS\,2 data, the spectra recorded in left- and right-hand polarized light, i.e.\ 
the ordinary and extraordinary beams
are extracted using a pipeline written in the MIDAS environment
by T.~Szeifert, the very first FORS instrument scientist.
This pipeline reduction by default includes background subtraction.
A single wavelength calibration frame is used for each night.

A first description of the assessment of the longitudinal magnetic field
measurements using FORS\,1/2 spectropolarimetric observations was presented 
in our previous work (e.g.\ Hubrig et al.\ \cite{Hubrig2004a,Hubrig2004b}, 
and references therein).

To minimize the cross-talk effect, a sequence of subexposures at the retarder
position angles $-$45$^{\circ}$+45$^{\circ}$, +45$^{\circ}$$-$45$^{\circ}$,
$-$45$^{\circ}$+45$^{\circ}$, etc.\ is usually executed during observations and
the $V/I$ spectrum is calculated using:

\begin{equation}
\frac{V}{I} = \frac{1}{2} \left\{ 
\left( \frac{f^{\rm o} - f^{\rm e}}{f^{\rm o} + f^{\rm e}} \right)_{-45^{\circ}} -
\left( \frac{f^{\rm o} - f^{\rm e}}{f^{\rm o} + f^{\rm e}} \right)_{+45^{\circ}} \right\}
\label{eq:1}
\end{equation}

where +45$^{\circ}$ and $-$45$^{\circ}$ indicate the position angle of the
retarder waveplate and $f^{\rm o}$ and $f^{\rm e}$ are the ordinary and
extraordinary beams, respectively.  Rectification of the $V/I$ spectra was
performed in the way described by Hubrig et al.\ (\cite{Hubrig2014}).
Null profiles, $N$, are calculated as pairwise differences from all available 
$V$ profiles.  From these, 3$\sigma$-outliers are identified and used to clip 
the $V$ profiles.  This removes spurious signals, which mostly come from cosmic
rays, and also reduces the noise. A full description of the updated data 
reduction and analysis will be presented in a separate paper (Sch\"oller et 
al., in preparation).

The mean longitudinal magnetic field, $\left< B_{\rm z}\right>$, is 
measured on the rectified and clipped spectra based on the relation
\begin{eqnarray} 
\frac{V}{I} = -\frac{g_{\rm eff}\, e \,\lambda^2}{4\pi\,m_{\rm e}\,c^2}\,
\frac{1}{I}\,\frac{{\rm d}I}{{\rm d}\lambda} \left<B_{\rm z}\right>\, ,
\label{eqn:vi}
\end{eqnarray} 

\noindent 
where $V$ is the Stokes parameter that measures the circular polarization, $I$
is the intensity in the unpolarized spectrum, $g_{\rm eff}$ is the effective
Land\'e factor, $e$ is the electron charge, $\lambda$ is the wavelength,
$m_{\rm e}$ is the electron mass, $c$ is the speed of light, 
${{\rm d}I/{\rm d}\lambda}$ is the wavelength derivative of Stokes~$I$, and 
$\left<B_{\rm z}\right>$ is the mean longitudinal (line-of-sight) magnetic field.

The magnetic field is usually measured in two ways: using only the absorption hydrogen Balmer
lines or using the entire spectrum including all available absorption lines in the spectra.
The Land\'e  factor $g_{\rm eff}$ is assumed to be 1.0 in the spectral regions containing hydrogen lines and
it is slightly larger (1.25) in the spectral regions outside the hydrogen lines.

The feasibility of longitudinal magnetic field measurements in massive stars 
using low-resolution spectropolarimetric observations was demonstrated by previous studies of O and B-type stars
(e.g., Hubrig et al.\ \cite{Hubrig2006,Hubrig2008,Hubrig2009,Hubrig2011a,Hubrig2011b,Hubrig2013a}).
The mean longitudinal magnetic field $\left<B_{\rm z}\right>$ is defined by the slope of the 
weighted linear regression line through the measured data points, where
the weight of each data point is given by the squared signal-to-noise ratio
of the Stokes~$V$ spectrum. The formal $1\,\sigma$ error of 
$\left<B_{\rm z}\right>$ is obtained from the standard relations for weighted 
linear regression. This error is inversely proportional to the rms  
signal-to-noise ratio of Stokes~$V$. Finally, the factor
$\sqrt{\chi^2_{\rm min}/\nu}$ is applied to the error determined from the 
linear regression, if larger than 1. No significant fields were detected in the null spectra 
calculated by combining the subexposures 
in such a way that polarization cancels out. This allows us to verify that no spurious signals are present in the data.

\begin{table}
\caption[]{
Magnetic field measurements of HD\,92207 using FORS\,2. 
All quoted errors are 1$\sigma$ uncertainties.
}
\label{tab:log_measfors}
\centering
\begin{tabular}{cr@{$\pm$}lr@{$\pm$}lr}
\hline \hline\\[-7pt]
\multicolumn{1}{c}{MJD} &
\multicolumn{2}{c}{$\left<B_{\rm z}\right>_{\rm all}$ [G]} &
\multicolumn{2}{c}{$\left<B_{\rm z}\right>_{\rm hyd}$ [G]} &
\multicolumn{1}{c}{S/N}\\
\hline\\[-7pt]
  56630.308& $-$64 & 93 &$-$132 & 238 & 680\\
  56672.104 &    104 & 34 &  231 & 148 & 2056\\
  56702.159 &     152 & 61 &  201 & 120 & 1675\\ 
\hline
\end{tabular}
\end{table}

\begin{figure}
\centering
\includegraphics[width=0.30\textwidth]{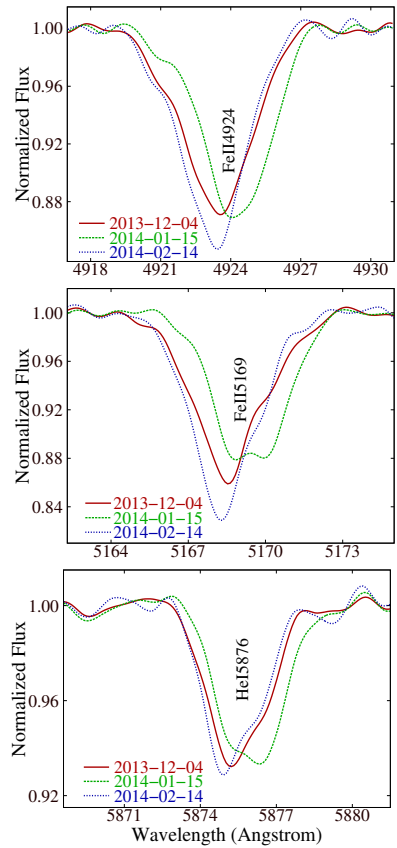}
\caption{
Variability of the Stokes~$I$ line profiles of  \ion{Fe}{ii} $\lambda$4924, \ion{Fe}{ii} $\lambda$5169, 
and \ion{He}{i} $\lambda$5876 (from top to bottom)
observed on three epochs in the FORS\,2 spectra. 
}
\label{fig:si}
\end{figure}

Furthermore, we have carried out Monte Carlo bootstrapping tests. 
These are most often applied with the purpose of deriving robust estimates of standard errors. 
In these tests, we randomly generate 250\,000 samples from the original data
that have the same size as the original data set and analyse 
the distribution of the  $\left<B_{\rm z}\right>$ determined from all these newly generated data sets. 
The measurement uncertainties obtained before and after Monte Carlo bootstrapping tests were found to be 
in close agreement, indicating the absence of reduction flaws. 
Distributions from our Monte Carlo bootstrapping tests for the data sets of
HD\,92207 from the three different epochs are presented in Fig.~\ref{fig:boot}.
The results of our magnetic field measurements along with the information related to the modified Julian date
and the signal-to-noise ratio in the observed spectra are listed in 
Table~\ref{tab:log_measfors}.
A magnetic field at a significance level of 3$\sigma$ was achieved 
during observations on 2014 January 15. We also find a change of polarity of the field between the observations 
obtained in 2013 December and 2014 January.

\subsection{Variability}
\label{sect:fors_var}

As has already been reported in the work by Hubrig et al.\ (\cite{Hubrig2014}), the previous analysis of FORS\,2 polarimetric 
spectra clearly showed the presence of short-term spectral 
variability on time scales of the order of minutes and less.
Along with radial velocity 
shifts up to 30\,km\,s$^{-1}$ detected for lines belonging to different elements, the authors also 
detected notable changes in line intensities up to 3\%.  Other stars observed during the same night,
also directly after the observations 
of HD\,92207 at the same air mass and with similar short exposure times did not exhibit a similar kind
of variability, indicating that spectral variability of HD\,92207 was not introduced by 
imperfection of the instrument. Hubrig et al.\ (\cite{Hubrig2014}) concluded that the detected spectral variability is 
intrinsic, but with the current data, it can not be decided, if the variations are of
periodic or stochastic nature. 

\begin{figure}
\centering
\includegraphics[width=0.46\textwidth]{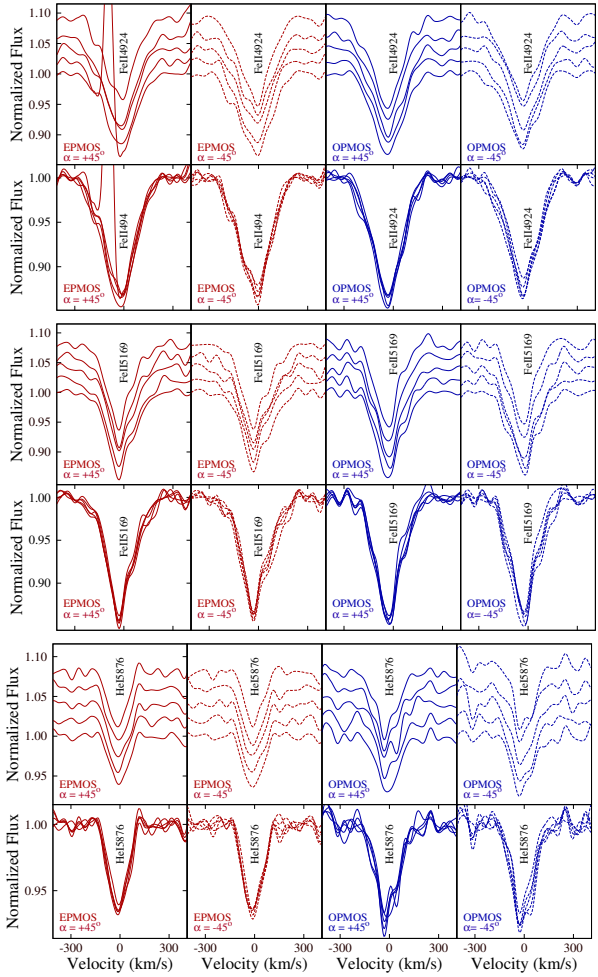}
\caption{The behaviour of the \ion{Fe}{ii} $\lambda$4924, \ion{Fe}{ii} $\lambda$5169, and 
\ion{He}{i} $\lambda$5876 lines (from top to bottom) in the ordinary and 
extraordinary beams of each individual subexposure recorded in 2013 December.
In the respective upper row, the line profiles are shifted in vertical
direction for best visibility, while overplotted line profiles are presented in the respective lower row.
Note that one \ion{Fe}{ii} $\lambda$4924 profile in the EPMOS sub-exposure is distorted by a cosmic.}
\label{fig:fe1}
\end{figure}

\begin{figure}
\centering
\includegraphics[width=0.46\textwidth]{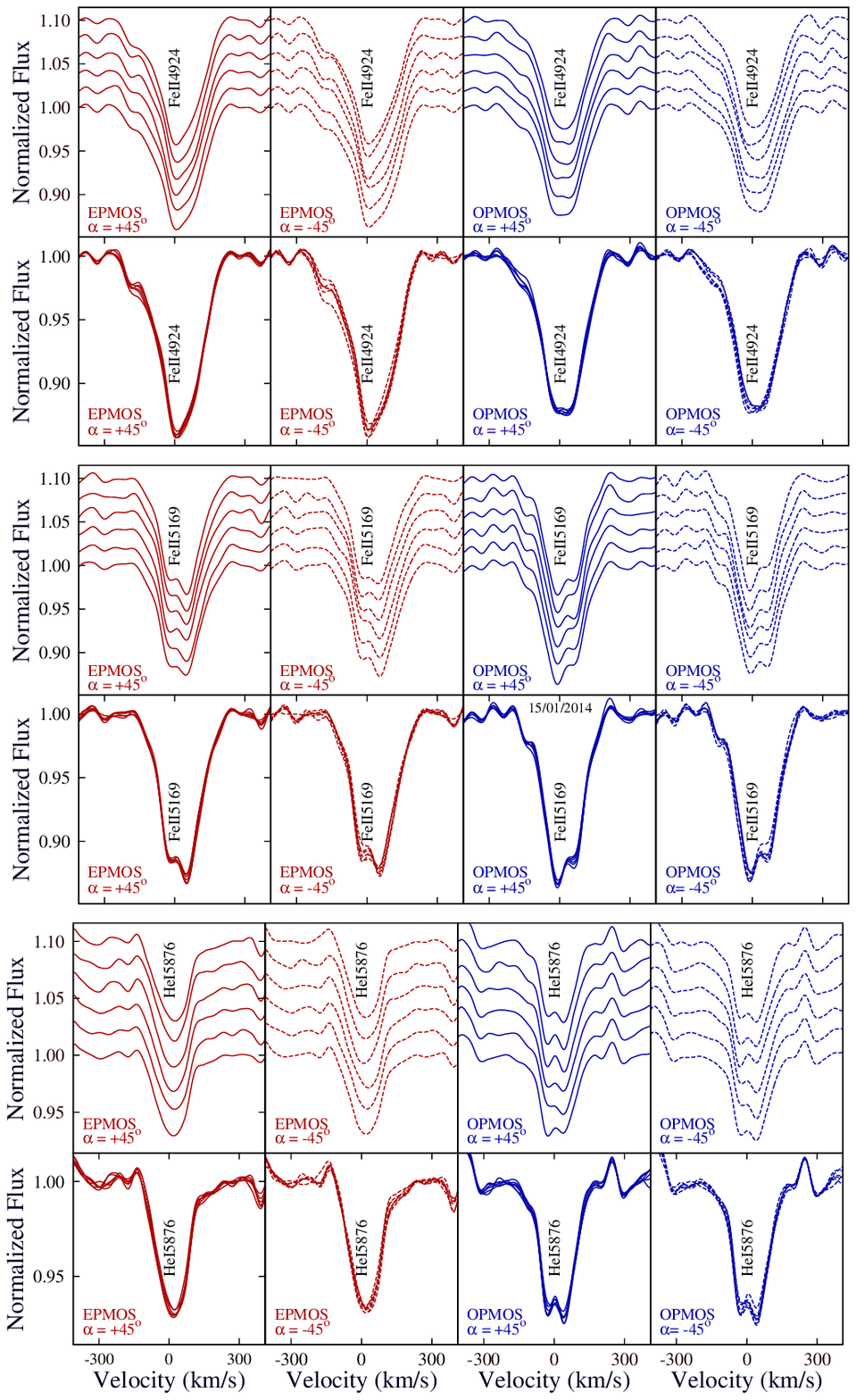}
\caption{ The behaviour of the \ion{Fe}{ii} $\lambda$4924, \ion{Fe}{ii} $\lambda$5169, and 
\ion{He}{i} $\lambda$5876 lines (from top to bottom) in the ordinary and 
extraordinary beams of each individual subexposure recorded in 2014 January.
In the respective upper row, the line profiles are shifted in vertical
direction for best visibility, while overplotted line profiles are presented in the respective lower row.
}
\label{fig:fe2}
\end{figure}

\begin{figure}
\centering
\includegraphics[width=0.46\textwidth]{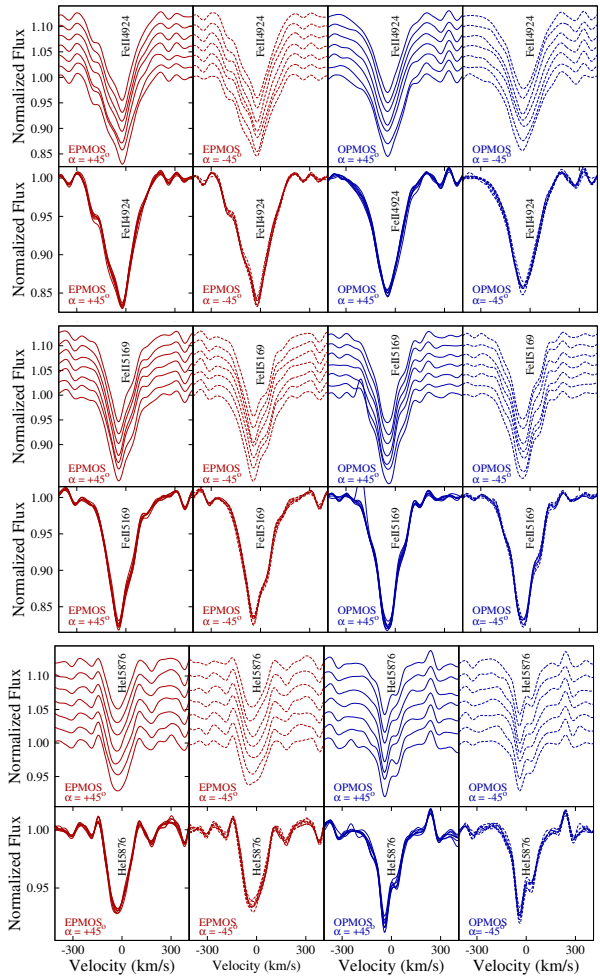}
\caption{The behaviour of the \ion{Fe}{ii} $\lambda$4924, \ion{Fe}{ii} $\lambda$5169, and 
\ion{He}{i} $\lambda$5876 lines (from top to bottom) in the ordinary and 
extraordinary beams of each individual subexposure recorded in 2014 February.
In the respective upper row, the line profiles are shifted in vertical
direction for best visibility, while overplotted line profiles are presented in the respective lower row.}
\label{fig:fe3}
\end{figure}

In our recent spectropolarimetric observations of HD\,92207 with FORS\,2, the Stokes~$I$ profile shape of spectral lines 
changes from one epoch to the next. The differences in the line behaviour are illustrated
in Fig.~\ref{fig:si}, where we display as an example the behaviour of the  
\ion{Fe}{ii} $\lambda$4924, \ion{Fe}{ii} $\lambda$5169, and \ion{He}{i} $\lambda$5876 lines with Land\'e factors
1.7, 1.3, and 1.0, respectively. An interesting fact 
observed in the spectra obtained in 2014 January, i.e.\ at the epoch when the longitudinal magnetic field is detected at a
significance level of 3$\sigma$, is that the lines \ion{Fe}{ii} $\lambda$5169 and 
\ion{He}{i} $\lambda$5876 appear slightly split. 
To try to understand the character of the spectral variability of HD\,92207 in FORS\,2 spectra 
on a short time scale and its impact on our
measurements of the longitudinal magnetic field, we decided to investigate the behaviour of the line profile
in left- and right-hand circular polarized light, i.e.\ in ordinary and extraordinary beams obtained at 
the retarder waveplate positions +45$^{\circ}$ and $-$45$^{\circ}$, respectively. Such a study appears reasonable 
since in the presence of 
a non-zero longitudinal magnetic field, due to the Zeeman effect,
the line profiles display a radial velocity shift or/and
a different profile shape.

As we already mentioned above, for spectropolarimetric observations with FORS\,2, a sequence of subexposures at the retarder
position angles $-$45$^{\circ}$+45$^{\circ}$, +45$^{\circ}$$-$45$^{\circ}$,
$-$45$^{\circ}$+45$^{\circ}$, etc., is usually executed during observations. Ten subexposures with an exposure 
time of 2\,sec were obtained in 2013 December, whereas twelve subexposures with an exposure 
time of 4\,sec were taken in 2014 January, and fourteen subexposures with an exposure 
time of 10\,sec were taken in 2014 February.
In Figs.~\ref{fig:fe1}--\ref{fig:fe3}, we present the behaviour of line profiles in the ordinary and 
extraordinary beams at each epoch.
In the presented figures, the ordinary and extraordinary beams are labeled as OPMOS and EPMOS, respectively. 
Please note that a spectrum from an ordinary beam at an angle of $-$45$^{\circ}$ corresponds to a spectrum
from an extraordinary beam at an angle of +45$^{\circ}$, and vice versa.
The differences seen on Figs.~\ref{fig:fe1}--\ref{fig:fe3} between the corresponding beams are due
to an imperfect wavelength calibration and are cancelling out in the $V/I$ spectrum, after the 
application of Eq.~\ref{eq:1}.

No significant variation between the line profile shapes in the ordinary and extraordinary beams are observed
in FORS\,2 polarimetric spectra obtained in 2013 December. 
At this epoch, no significant longitudinal magnetic field was detected. 
In contrast, in 2014 January, the line profiles 
in opposite circularly polarized light exhibit remarkable differences:
the line profiles of all three lines in the ordinary beams
appear slightly split. The line  \ion{Fe}{ii} $\lambda$5169 exhibits split profiles also in the extraordinary beam, 
but the shape of the line profile is different compared to that observed in the ordinary beams.
At that epoch, we measure a longitudinal magnetic field at a significance level of 3$\sigma$.
In 2014 February, the profiles of the \ion{Fe}{ii} $\lambda$4924 and \ion{Fe}{ii} $\lambda$5169 lines in
both ordinary and extraordinary beams appear similar, and  the \ion{Fe}{ii} $\lambda$5169 line profile 
shows a small distortion in the red wing.
However, the behaviour of the \ion{He}{i} $\lambda$5876 line is different, indicating the presence of a small splitting
in the profile shape observed in the ordinary beams.
At this epoch, a longitudinal magnetic field is measured at a
significance level of 2.5$\sigma$. 

The detected behaviour of the line profile in opposite circularly polarized light has not been reported before. The inspection of the  
previous FORS\,2 spectropolarimetric observations discussed in the work by Hubrig et al.\ (\cite{Hubrig2014}) revealed that also 
in the observations obtained at MJD 55936.341 the lines  
\ion{Fe}{ii} $\lambda$4924 and \ion{He}{i} $\lambda$5876 show a similar splitting in both ordinary and extraordinary beams, while
the line \ion{Fe}{ii} $\lambda$5169 does not show any splitting. No splitting in any line is detected on the dates 
MJD55688.168 and MJD56018.223.
Since apart from the three lines discussed above, only very few other 
\ion{Fe}{ii} lines show similar split line profile shapes, we can conclude that the detected splitting is not a result of an imperfect data
reduction. The interesting fact is that the lines \ion{Fe}{ii} $\lambda$4924 and \ion{Fe}{ii} $\lambda$5169 belong to 
multiplet 42.
Lines of this multiplet present the principle emission lines observed in individual classes of emission objects, among them 
Oe-, Be-, and P\,Cygni-stars.
However, the third member of this multiplet, the line \ion{Fe}{ii} $\lambda$5018,
was not found split in any FORS\,2 spectropolarimetric observation.

We have currently no explanation for the appearance 
of the line splitting observed in the polarized line profiles.
We note that since previously only very few  searches for the 
presence of magnetic fields in a small number of early A-type supergiants were conducted, mostly using 
least-squares deconvolution i.e.\ without any presentation of 
individual line profiles in opposite circular polarization (e.g., Verdugo et al.\ \cite{Verdugo2005}), the necessary expertise for any 
well-founded explanation of the observed behaviour is missing.
On the other hand, given the size and the shape of the splitting, it is obvious that the splitting is not related to 
the Zeeman splitting frequently observed in slowly rotating  Ap and Bp stars
with strong magnetic fields.
It is also likely that such a splitting can not be related to the appearance 
of emission in the line cores since the observed H$\beta$ line in the FORS\,2 spectra does not appear in emission 
on any of the three epochs. Noteworthy, we are not able to inspect the behaviour of the H$\alpha$ line in our spectra,
as it is located outside of the spectral range available with our FORS\,2 setting. Further, it is also not clear whether any
kind of pulsation would affect polarized profiles causing the appearance of the line splitting in polarized light.

\begin{figure*}
\centering
\includegraphics[width=0.46\textwidth]{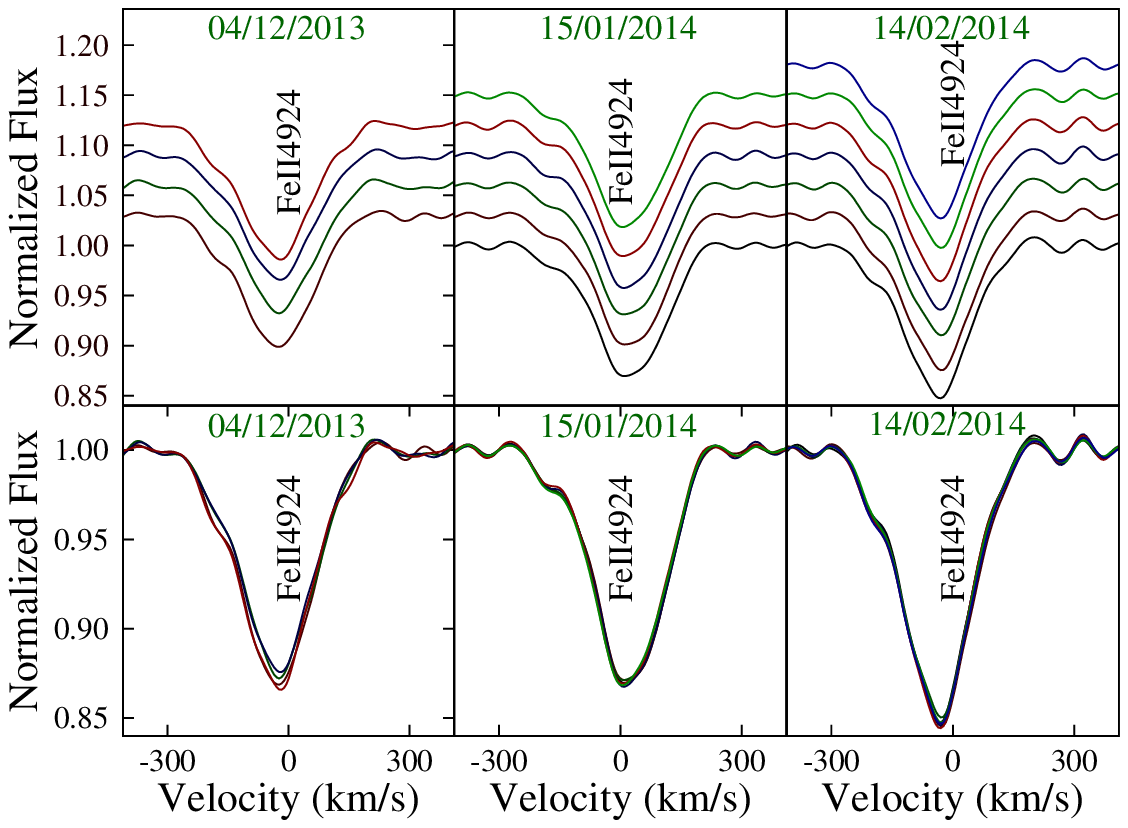}
\includegraphics[width=0.46\textwidth]{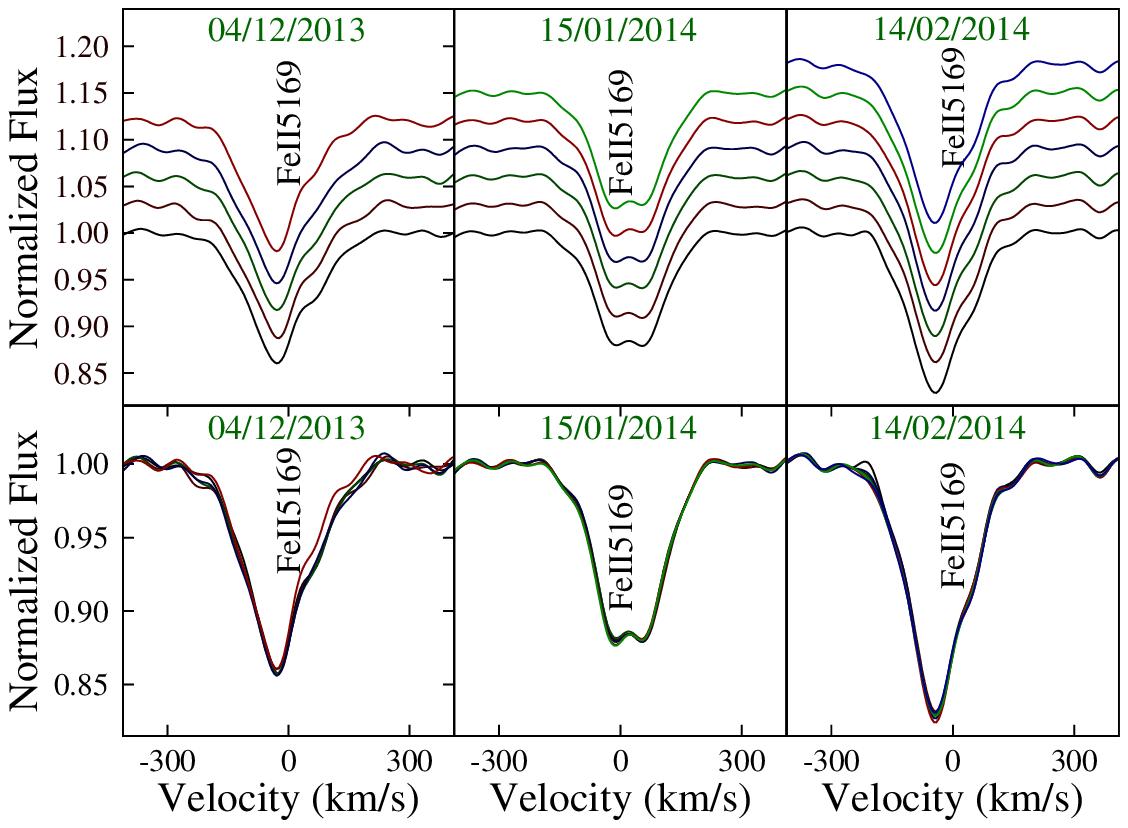}
\caption{
FORS\,2 Stokes~$I$ profiles of the \ion{Fe}{ii}\,$\lambda$4924 (left panel) and 
\ion{Fe}{ii}\,$\lambda$5169 (right panel) lines from individual subexposures of all three epochs. 
For each epoch, in the upper rows, we present the line profiles shifted in vertical direction for best visibility.
The lower rows show all profiles overplotted.}
\label{fig:stokesi}
\end{figure*}

Hubrig et al.\ (\cite{Hubrig2014}) reported the presence of radial velocity 
shifts of up to 30\,km\,s$^{-1}$ and changes in line intensities of up to 3\%.
Variations of the line profiles in Stokes~$I$ spectra with radial velocities reaching 10--20\,km\,s$^{-1}$ are observed 
in the line profiles 
recorded in 2013 December and 2014 February, while during the observations in 2014 January the line profiles 
appear significantly stable with measured radial velocity shifts below 3\,km\,s$^{-1}$.
In all observed spectra is
the intensity variation below $\sim$2\% and mostly caused by a stronger noise in the line cores. 
These results indicate that the short-term spectral variability most likely does not have any significant impact
on our magnetic field measurements using the spectra obtained in 2014 January.
As an example, we present in Fig.~\ref{fig:stokesi} overplotted Stokes~$I$ profiles of the \ion{Fe}{ii}\,$\lambda$4924 and 
\ion{Fe}{ii}\,$\lambda$5169 lines in each individual subexposure on all three epochs. The shifts of individual line 
profiles appear in general substantially smaller in 2014 January than those presented by 
Hubrig et al.\ (\cite{Hubrig2014}) in their Figs.~2--5.

\section{Measurements of the magnetic field using HARPS observations}
\label{sect:harps}

\begin{table}
\caption[]{
Magnetic field measurements of HD\,92207 using HARPS polarimetric spectra.
All quoted errors are 1$\sigma$ uncertainties.
}
\label{tab:log_meas}
\centering
\begin{tabular}{ccr@{$\pm$}lr@{$\pm$}l}
\hline \hline\\[-7pt]
\multicolumn{1}{c}{HJD} &
\multicolumn{1}{c}{S/N} &
\multicolumn{2}{c}{$\left<B_{\rm z}\right>_{\rm Fe}$ [G]} &
\multicolumn{2}{c}{$\left<B_{\rm z}\right>_{\rm Fe,n}$ [G]} \\
\hline\\[-7pt]
 2456345.677    &345  &$-$36 & 21 & $-$12 &25\\
 2456347.676    &510  &  11 & 11 &  5 & 15 \\
 2456348.677    &450  &2 & 13 &  $-$3  & 16 \\
 2456350.669    &485  &14 & 11 &  6  & 14 \\
\hline\\
\end{tabular}
\end{table}

Part of the spectra analyzed in this work were obtained with the HARPS polarimeter 
(Snik et al.\ \cite{snik2008}) feeding the HARPS spectrometer 
at the ESO 3.6-m telescope on La Silla. 
We downloaded from the ESO archive the publically available polarimetric spectra of HD\,92207, obtained on four 
different epochs in 2013 February. 
All spectra have a resolving power of $R = 115\,000$ and the signal-to-noise ratio $(S/N)$ in the Stokes~$I$
profiles is between 345 and 510.
The HARPS archival spectra cover the wavelength range 3780--6913\,\AA{}, with a small gap around 5300\,\AA{}. 
Each observation of the star is split into four sub-exposures, 
obtained with four different orientations of the quarter-wave retarder plate relative to the 
beam splitter of the polarimeter.
The reduction was performed using the HARPS data reduction 
software available at the ESO headquarters in Garching. The normalization procedure for the HARPS spectra to the 
continuum level is described in detail in the work of Hubrig et al.\ (\cite{Hubrig2013b}).
The Stokes~$I$ and $V$ parameters were derived following the ratio method described by 
Donati et al.\ (\cite{Donati1997}), 
ensuring in particular that all spurious signatures are removed up to first order. 
The information on the individual observations, including the dates of observation and the $S/N$ values, is 
given in Table~\ref{tab:log_meas}.

\begin{figure*}
\centering
\includegraphics[angle=0,width=0.58\textwidth]{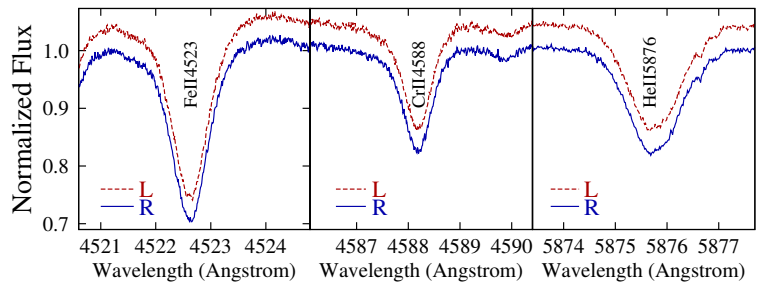}
\caption{
The line profiles of different spectral lines, averaged over the four sub-exposures, recorded with HARPS in 
left- and right-hand polarized light (indicated by the letters $L$ and $R$) on 2013 February 25.
The spectrum recorded in right-hand polarized light is shifted in vertical 
direction for better visibility. 
}
\label{fig:hd94}
\end{figure*}

We used for the measurements of the longitudinal magnetic field
the moment technique developed by Mathys (e.g.\ \cite{Mathys1991}).
The diagnostic potential of high-resolution circularly polarized spectra using the moment technique has been 
discussed at length in numerous papers by Mathys (e.g.,\ \cite{Mathys1993}, \cite{Mathys1994}).
Wavelength shifts between right- and left-hand side circularly polarized spectra are
interpreted in terms of a longitudinal magnetic field $\left<B_{\rm z}\right>$.
Furthermore, this technique allows us not only the determination of the mean longitudinal magnetic field, but 
also to investigate the presence of crossover effect, and to measure the quadratic magnetic field.
For each line in the selected sample of 50 \ion{Fe}{ii} lines, the measured shifts between the line profiles in the 
left- and right-hand circularly polarized HARPS spectra
are used in a linear regression analysis in the
$\Delta\lambda$ versus $\lambda^2 g_{\rm eff}$ diagram, following the formalism discussed by 
Mathys (\cite{Mathys1991,Mathys1994}). 
The measured values for the mean longitudinal magnetic field $\left< B_{\rm z} \right>$ are presented in 
Table~\ref{tab:log_meas}. They range from $-$36\,G obtained on the first
observing night to $+$14\,G obtained on the last observing night.  
We do not detect a longitudinal magnetic field at a significance level of 3$\sigma$ for any night,
and, as we show below, on each night the profile shapes of spectral lines recorded in each subexposure in 
the left- and right-hand polarized light appear rather identical, indicating the absence of a significant 
longitudinal magnetic field.
As an example, we present in Fig.~\ref{fig:hd94} the line profiles recorded with HARPS in left- and 
right-hand polarized light on 2013 February 25.

As a crosscheck, we measure the mean longitudinal magnetic field
on null polarization spectra, which are calculated by combining the subexposures 
in such a way that polarization cancels out. The results of our measurements are presented in 
Table~\ref{tab:log_meas}. The measurements on the spectral lines of \ion{Fe}{ii} 
using null spectra are labeled with $n$.
Since no significant fields could be determined from null spectra, we conclude that
any noticeable spurious polarization is absent. 
No significant crossover or mean quadratic magnetic field has been detected on the four observing epochs. 

\begin{figure}
\centering
\includegraphics[angle=0,width=0.39\textwidth]{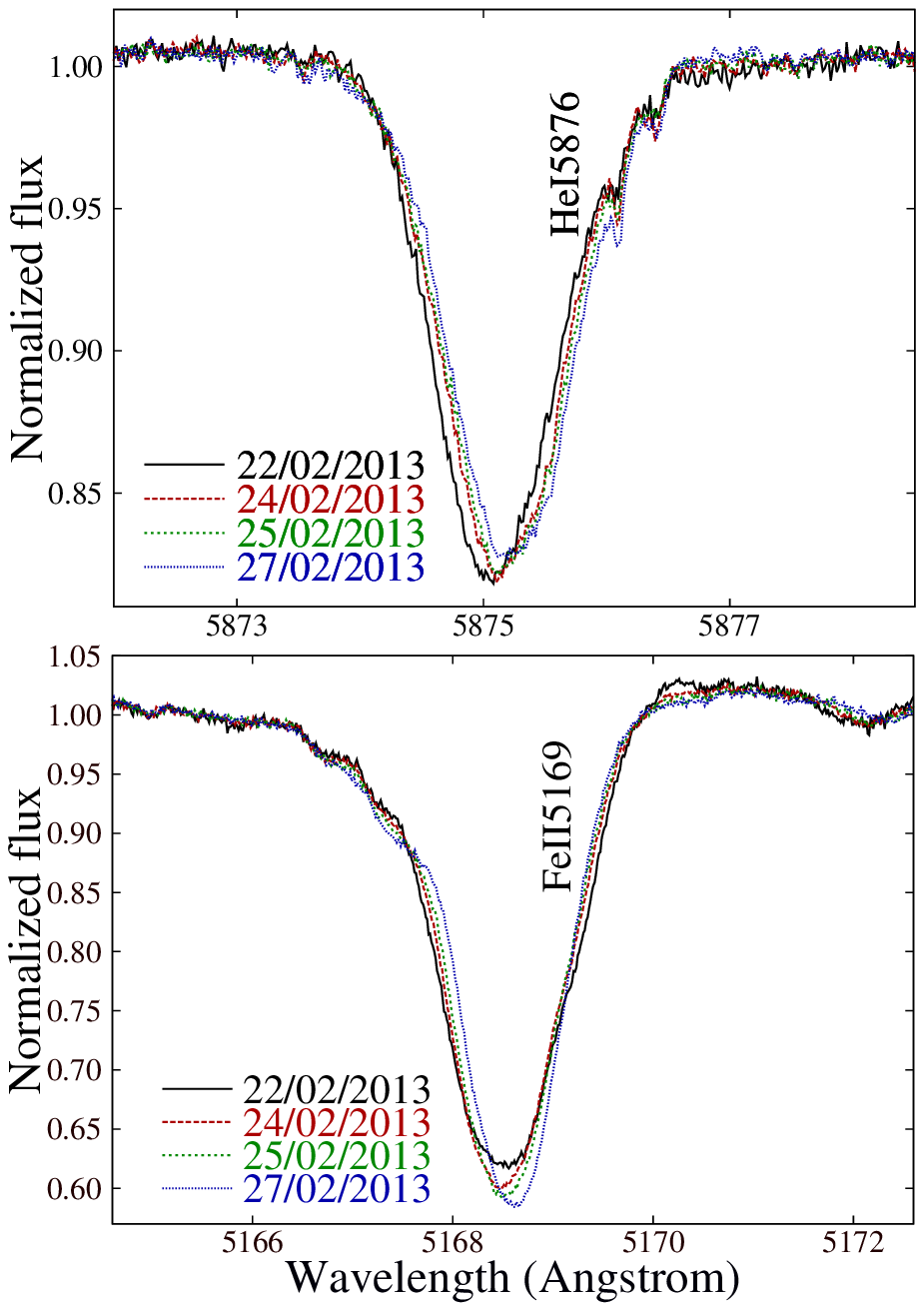}
\caption{
{\sl Upper panel:} The variability of the \ion{He}{i}\,$\lambda$5876 line in the HARPS spectra 
of HD\,92207 observed from 2013 February 22 to 27.
{\sl Lower panel:}  The variability of the \ion{Fe}{ii}\,$\lambda$5169 line over the same nights.
}
\label{fig:harps4}
\end{figure}

Clear day-to-day spectral variations are discovered in the spectra observed  from 2013 February 22 to 27.
We detect that the radial velocity for metallic lines is increasing from night to night at a mean rate of  
1.7\,km\,s$^{-1}$, while for the \ion{He}{i} lines it increases by 0.8\,km\,s$^{-1}$. The intensities of metallic lines vary opposite
to the intensities of \ion{He}{i} lines. We observe over the five nights 
the increase of the metallic line intensity by about 3\% while the intensity for the \ion{He}{i} lines decreases by about 1\%.
As an example of the day-to-day spectral variations, we present in Fig.~\ref{fig:harps4} the behaviour of the 
\ion{Fe}{ii}\,$\lambda$5169 and \ion{He}{i}\,$\lambda$5876 lines.

\begin{figure*}
\centering
\includegraphics[angle=0,width=0.75\textwidth]{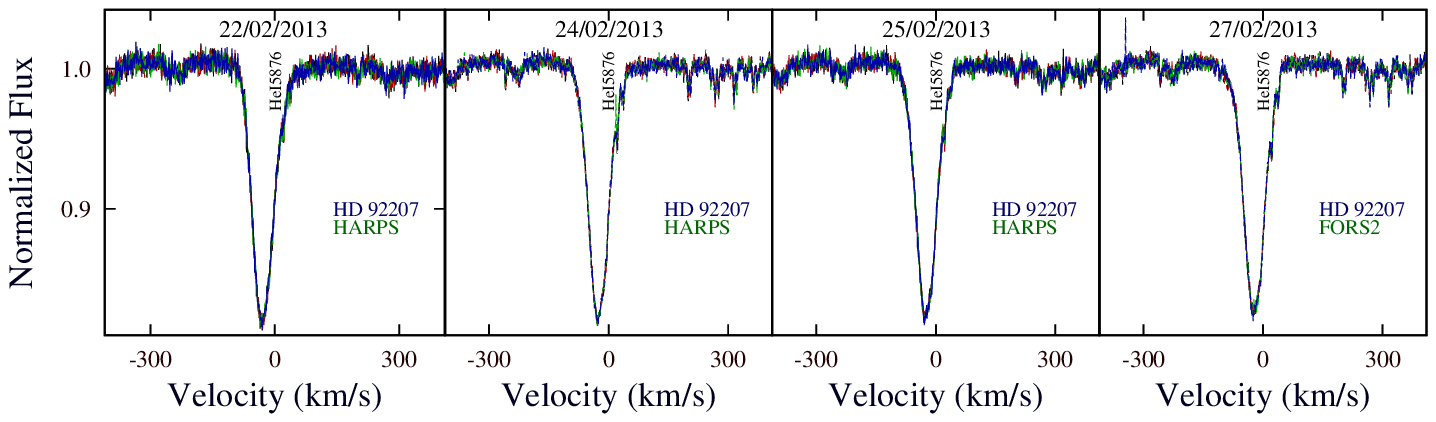}
\includegraphics[angle=0,width=0.75\textwidth]{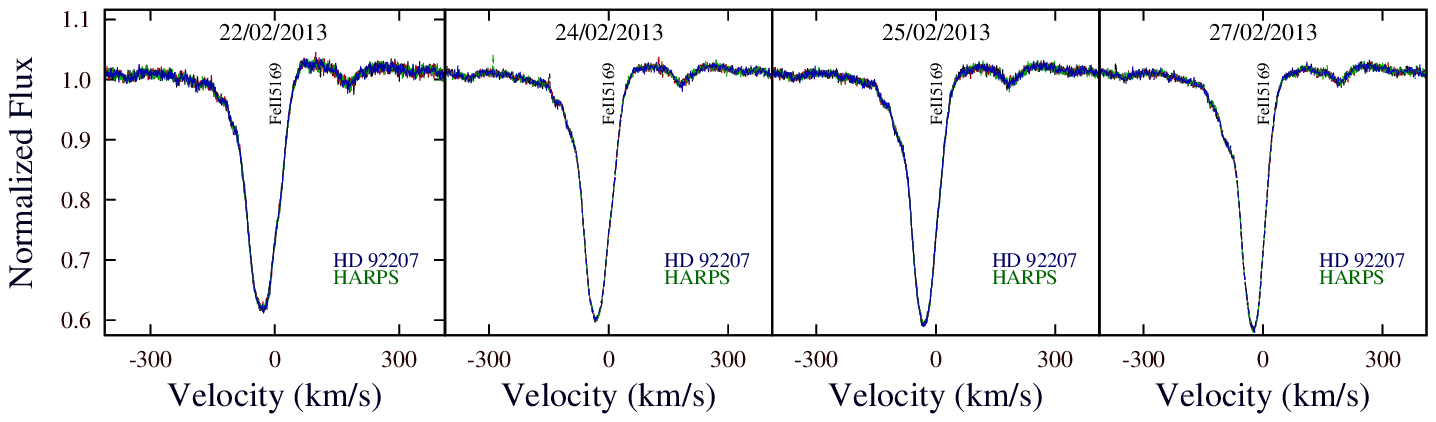}
\caption{
{\sl Upper panel:} The behaviour of the \ion{He}{i}\,$\lambda$5876 line in the HARPS spectra 
of HD\,92207 over four nights from 2013 February 22 to 27 (from left to right).
Each panel presents the overplotted four Stokes~$I$ 
spectra obtained over each night.   
{\sl Lower panel:} The behaviour of the \ion{Fe}{ii}\,$\lambda$5169 line over the same nights.  
}
\label{fig:hd93}
\end{figure*}

Since HARPS spectropolarimetric observations are usually split into four sub-exposures, it is also possible to test
the line profile variability not only on the day-to-day time scale but also on shorter time scales corresponding
to the duration of sub-exposures, which in the case of HD\,92207 accounts for 8.3 to 10\,min. 
Admittedly, these exposure times are much longer than those used for FORS\,2 observations. On the other hand, the 
analysis of HARPS spectra allows us to explore the presence of the variability on 8--10 minute time scales.
In Fig.~\ref{fig:hd93},
we present the profiles of the  \ion{Fe}{ii}\,$\lambda$5169 and \ion{He}{i}\,$\lambda$5876 lines recorded 
on each individual night. No obvious changes in radial velocity or intensity are detected at these time scales.
In the sub-exposures obtained during the same night, the flux differences
are of the order of 0.5\% of the continuum level, which is consistent with the noise estimate. 

\section{Spectroscopic time-series obtained with the CORALIE spectrograph}
\label{sect:CORALIE}

\begin{figure}
\centering
\includegraphics[width=0.35\textwidth]{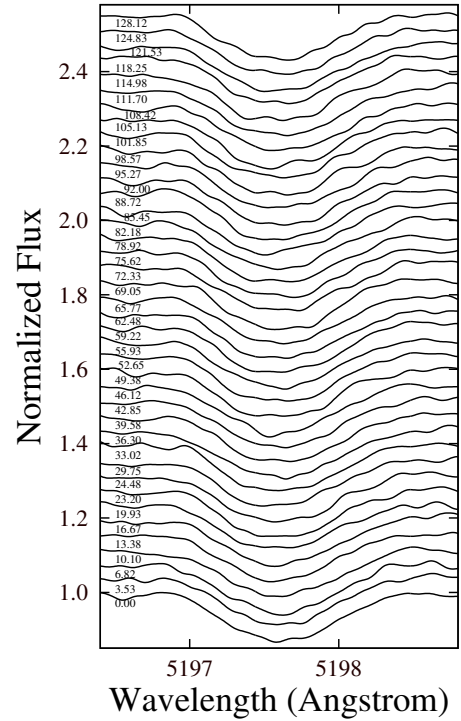}
\caption{
Low variability of the line profile of \ion{Fe}{ii}\,$\lambda$5197.6 in the CORALIE spectra observed on 2014 January 8.
}
\label{fig:var}
\end{figure}

To search for the presence of variability in HD\,92207 in high-resolution spectra on even shorter time scales of 
the order of a few minutes, 
we used the CORALIE spectrograph installed at 
the Swiss 1.2-m Leonard Euler telescope on La Silla. CORALIE is a fibre-fed echelle spectrograph with $R\sim60\,000$
and with a wavelength coverage from 3800 to 6900\,\AA{}.
The instrument is described by Queloz et al. (\cite{que2001}), and instrumental upgrades made in 2007 are 
detailed by S\'egransan et al. (\cite{seg2010}). The raw data were reduced using the efficient automated reduction 
pipeline, which follows standard procedures and performs pre- and overscan bias correction, flatfielding using 
Halogen lamps, background modelization, as well as cosmic removal. ThAr lamps are used for wavelength calibration.

Two time-series over several hours were obtained on the nights in 2014 January 8 and 12.
During the first night, we were able to record 40 spectra with an exposure time of 120\,s and a signal-to-noise ratio 
of about 80--100.
Taking into account the readout overheads, the individual exposures are separated by 3.28\,min.
In Fig.~\ref{fig:var}, we present the behaviour of the line profile of the \ion{Fe}{ii}\,$\lambda$5197.6 line in 
all 40 spectra.
Due to the presence of the rather strong noise, before plotting the line profiles, we have slightly smoothed 
the profiles
using a Gauss filter with a width of 0.05\,\AA{}. Although the line profiles appear to be slightly variable,
the level of variability is rather low, much less than that observed by 
Hubrig et al.\ (\cite{Hubrig2014}) in FORS\,2 spectra.  

During the second night on January 12, an additional time series of 29 and 28 spectra with an exposure time 
between 120 and 150\,sec and similar S/N, separated by a wavelength calibration exposure, were recorded.
The total observing time accounted for 3\,h 40\,min.
In these spectra, one can find similar line profile variations as in the night of January 8.

\begin{figure}
\centering
\includegraphics[width=0.48\textwidth]{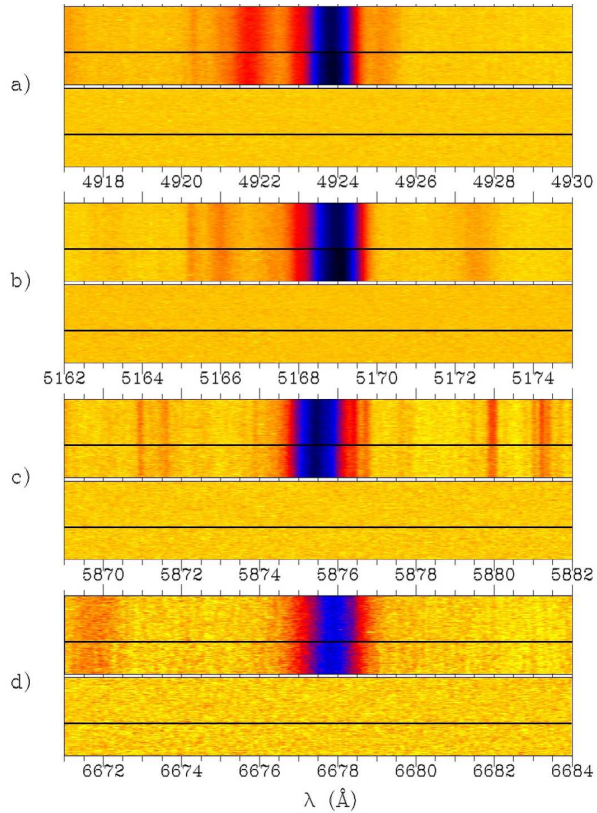}
\caption{Stacked CORALIE spectra of HD\,92207 with subtracted mean spectrum for the first and the second series in the 
spectral regions around
a) \ion{Fe}{ii}\,$\lambda$4924,
b) \ion{Fe}{ii}\,$\lambda$5169, 
c) \ion{He}{i}\,$\lambda$5876, and
d) \ion{Fe}{i}\,$\lambda$6678.
The upper panels present the observed spectra and the lower panels the residual spectra
with respect to the mean spectrum.
}
\label{fig:stacked_rms}
\end{figure}

To check that the observed variability is not just due to the impact of the noise in the observed spectra, 
we constructed 
images of stacked spectra in several spectral regions.
In Fig.~\ref{fig:stacked_rms} in the upper panels, we present the stacked CORALIE spectra for the first and the 
second series in different spectral regions around the lines \ion{Fe}{ii}\,$\lambda$4924, \ion{Fe}{ii}\,$\lambda$5169,
\ion{He}{i}\,$\lambda$5876, and \ion{Fe}{i}\,$\lambda$6678.
In the lower panels we show the residual spectra with respect to the mean spectrum.
All six panels are 13\,\AA{} long, and in all presented images, the $y$-dimension corresponds to the observing time.
The inspection of these images seems to indicate that some structure is present in the cores of 
the \ion{He}{i}\,$\lambda$5876 and \ion{Fe}{i}\,$\lambda$6677 lines.
However, as we show in the same figure in the lower panels, if we subtract the mean 
spectrum calculated using all observed spectra, this structure disappears, indicating
that the amplitude of the structure is of the same order as the spectrum noise.
The presented results can be explained in two different ways:  either short-term spectral variability 
on a time scale of 3-4\,min does not exist, or the amplitude of the line profile variability is of the same order as 
the spectrum noise and cannot be detected in spectra obtained with a signal-to-noise ratio of about 80--100.

\section{Discussion}
\label{sect:disc}

We obtained new magnetic field measurements of HD\,92207 on three different epochs in 2013 and 2014 using FORS\,2 in 
spectropolarimetric mode.
Previous FORS\,2 observations revealed the presence of a weak mean longitudinal
magnetic field, but the definite 
confirmation of the magnetic nature of this object was pending due to the detection of short-term spectral 
variability probably
affecting the position of line profiles in left- and right-hand polarized spectra. 
In our recent  FORS\,2 observations obtained in 2014 January, the position of spectral lines appeared stable, with 
measured radial velocity shifts below 3\,km\,s$^{-1}$.
Exactly at this epoch,  using the entire spectrum 
including all hydrogen lines for the measurement of the magnetic field, we achieved a 3$\sigma$ detection
with $\left<B_{\rm z}\right>_{\rm all}=104\pm34$\,G. This result is fully in agreement with the previous detection
of a weak longitudinal magnetic field in this star.

However, our analysis of the polarimetric FORS\,2 spectra revealed a strange phenomenon that is currently difficult to explain
and that has never been mentioned by other polarimetric studies of blue supergiants before: we discover that the line profiles in the light 
polarized in a certain direction appear slightly split.
Given the size and the shape of the splitting, it is obvious that the splitting is not related to 
the Zeeman splitting frequently observed in slowly rotating  Ap and Bp stars
with strong magnetic fields.
We also cannot ascribe this splitting to the appearance of an emission in the line core, since 
the H$\beta$ line, which is sensitive to the stellar wind, does not display any core emission or splitting in
polarized light.
Further, we are also not aware of any kind of pulsations that would affect polarized profiles in such a way.
Clearly, future monitoring of the behaviour of the line profiles in polarimetric spectra will be very valuable
to identify the mechanism causing the line splitting in circularly polarized light.

The available high-resolution polarimetric HARPS spectra on four different epochs show clear day-to-day variations in 
all spectral lines, but short-term variability on a time scale of 8--10\,min is not detected.
We do not find any distinct Zeeman features
in the Stokes~$V$ spectra and the determined value of the longitudinal magnetic field is always below the 3$\sigma$ level.
No significant crossover or mean quadratic magnetic field is detected in the four HARPS  observations 

To search for short-term variability on even shorter timescales down to 3--4\,min, we obtained time series at two 
different epochs with the CORALIE spectrograph.
The structure visible in a few spectral lines in the images presenting stacked spectra disappears if the mean spectrum is 
subtracted. Due to the rather low signal-to-noise ratio of CORALIE spectra, we can not rule out that 
spectral variability is present, but its amplitude is of the same order as 
the spectrum noise. 

Despite several decades of observational
efforts with ground-based photometry and spectroscopy of bright blue supergiants,
it is not yet certain what portion of
their variability is periodic, or how far they deviate from strict periodicity. 
As a matter of fact, short-term variability was already identified
on a time scale of 1--3\,hours (Lefever et al.\ \cite{Lefever2007}; Kraus et al.\ \cite{Kraus2012}). 
However, a variability on time scales of the order of minutes and less has not been investigated so far.
To improve the sensitivity to such variability, it would be important to use for the time series a high-resolution 
spectrograph installed
at a telescope with a larger collecting area, such as the UV-Visual Echelle Spectrograph attached to the VLT 8\,m 
Kuyen telescope.


\label{lastpage}

\end{document}